\documentclass[a4paper,11pt]{article}
\pdfoutput=1 

\usepackage{jheppub} 

\usepackage[T1]{fontenc} 
\usepackage{comment}
\begin{document}
\title{\boldmath The Cosmic Neutrino Background on the Surface of the Earth}


\author[a]{A. Arvanitaki}
\author[a,b]{and S. Dimopoulos}


\affiliation[a]{Perimeter Institute for Theoretical Physics, Waterloo, ON, Canada}
\affiliation[b]{Stanford University, Stanford, CA, USA}

\emailAdd{aarvanitaki@perimeterinstitute.ca}
\emailAdd{savas@stanford.edu}

\abstract{ We argue that the reflection of relic neutrinos from the surface of the Earth results in a significant local $\nu-\bar{\nu}$ asymmetry, far exceeding the expected primordial lepton asymmetry. 
The net fractional electron neutrino number $\frac{n_{\nu_e}-n_{\bar{\nu}_e}}{n_{\nu_e}}$ is up to $\mathcal{O}(10^5) \sqrt{\frac{m_\nu}{0.1~\text{eV}}}$ larger than that implied by the baryon asymmetry. This enhancement is due to the weak 4-Fermi repulsion of the $\nu_e$ from ordinary matter which  slows down the $\nu_e$ near the Earth's surface, and to the resulting evanescent neutrino wave that penetrates  below the surface. This repulsion thus creates a net $\nu_e$ overdensity in a shell $\sim 7~\text{meters} \sqrt{\frac{0.1~\text{eV}}{m_\nu}}$ thick around the Earth's surface. Similarly the repulsion between $\bar{\nu}_\mu$ or $\bar{\nu}_\tau$ and ordinary matter creates an overdensity of $\bar{\nu}_{\mu, \tau}$ of similar size. These local enhancements increase the size of $\mathcal{O}(G_F)$ torques of the $C\nu B$ on spin-polarized matter by a factor of order $10^5$.  In addition,  they create a gradient of the net neutrino density which naturally provides a way out of the forty-year-old ``no-go'' theorems on the vanishing of $\mathcal{O}(G_F)$ forces. The torque resulting from such a gradient force can be $10^8$ times larger than that of earlier proposals.
Although the size of these effects is still far from current reach, they may point to new directions for $C\nu B$ detection. 

}

\maketitle
\flushbottom

\section{Introduction}
\label{sec:intro}

The Cosmic Neutrino Background ($C\nu B$) originates from the pre-BBN era and encodes important information about the early universe and the neutrino flavor sector. If discovered, it extends our knowledge of the Universe to well before the time of the CMB. Following a massless Fermi-Dirac distribution, the $C\nu B$ has a temperature of $T_\nu=1.68\times 10^{-4}$~eV.

The weak interaction cross-section of these relic neutrinos with ordinary matter is extremely small, since it is proportional to $G_F^2$. In 1962, Weinberg pointed out~\cite{Weinberg:1962zza} that this cross-section can be enhanced when the relic neutrinos are absorbed in radioactive processes. This is the basis of the PTOLEMY proposal~\cite{PTOLEMY:2018jst}, which suggests looking for the absorption of the relic neutrinos by a tritium nucleus which subsequently converts to helium. The neutrino cross-section can also be enhanced through coherent scattering \cite{Shvartsman:1982sn, gelmininussinov:2001hd, Domcke:2017aqj, Shergold:2021evs, Bauer:2022lri}.


In addition to scattering, the weak 4-Fermi interaction of these relic neutrinos is responsible for refractive effects. Since the deBroglie wavelength of cosmic neutrinos, $\lambda_{dB}$, is much larger than the interatomic distance, the neutrinos are subject to an effective in-matter potential, $U$:
\begin{eqnarray}
\label{eq:U}
U=\frac{G_F}{2 \sqrt{2}}\rho_{\text{matter}}\times 
\begin{cases}
    (-)(3Z-A) & \text{for} ~\nu_e ~(\bar{\nu_e})\\
    (-)(Z-A) &  \text{for} ~\nu_{\mu,\tau} ~(\bar{\nu}_{\mu,\tau}),
\end{cases}
\end{eqnarray}
where $G_F$ is Fermi's constant, and $\rho_{\text{matter}}$ is the number density of atoms in the material. $U$ produces an index of refraction $n_\nu$ for neutrinos that is differs from one by an amount $\delta_\nu$~\cite{Langacker:1982ih}:
\begin{eqnarray}
n_\nu-1\equiv \delta_\nu=-\langle\frac{m_\nu U}{ k_\nu^2}\rangle.
\end{eqnarray} 

In the 70s, several proposals were put forth to look for a force or a torque on a macroscopic object due to these refractive effects~\cite{Lewis:1979mu,Opher:1974drq,Stodolsky:1974aq}. Unfortunately, Langacker et. al.~\cite{Langacker:1982ih}, as well as Cabibbo and Maiani~\cite{Cabibbo:1982bb} independently, showed that most of these effects are zero in a uniform $C\nu B$. The only proposal that survived their now famous ``no-go'' theorem for refractive effects of the $C \nu B$ is the one by Stodolsky \cite{Stodolsky:1974aq}. Stodolsky pointed out that any electron or nuclear spin will experience a torque in the $C \nu  B$ which is a equivalent to an energy difference between spin-up and spin-down states of $\frac{G_F}{2 \sqrt{2}} (n_\nu- n_{\bar{\nu}}) \vec{\upsilon}_{\text{rel}}\cdot \vec{\sigma}$. In addition to being suppressed by the relative velocity of the solar system to the $C \nu B$, there is a suppression due to the neutrino-antineutrino asymmetry which is naively of order $4.4 \times 10^{-9}$, the same as the baryon asymmetry. 

In this paper, we argue that the Earth is responsible for a local enhancement of the neutrino-antineutrino asymmetry as well as the generation of a gradient of the neutrino density close to the Earth's surface. One can see from eq.~\ref{eq:U} that electron neutrinos and muon and tau antineutrinos, for which $\delta_\nu<0$, will experience a repulsive potential as they intercept the Earth. $\nu_e$'s and $\bar{\nu}_{\mu,\tau}$'s whose momentum perpendicular to the surface of the Earth, $k_\perp$, is not large enough to overcome this repulsion, i.e:
\begin{eqnarray}
k_\perp\leq \sqrt{2 m_\nu U}\equiv k_{\perp_{\text{cr}}}, 
\end{eqnarray}
will be reflected from the surface. As they are reflected, they slow down and cause a local overdensity that counts the fraction of those reflected neutrinos, which is:
\begin{eqnarray}
\sqrt{\langle\frac{2 m_\nu U}{ k_\nu^2 }\rangle}= \sqrt{-2\delta_\nu}\equiv \theta_\text{cr}
\end{eqnarray}

where $\theta_\text{cr}$ is the critical angle for total reflection, and is proportional to $\sqrt{G_F}$. Numerically, as seen from the values of $\delta_\nu$ shown in table~\ref{tab:deltanu}, this fractional overdensity is $2.2 \times 10^{-4} \sqrt{\frac{m_\nu}{0.1~\text{eV}}}$, which is several orders of magnitude bigger than the naive expectation of $\frac{n_{\nu}-n_{\bar{\nu}}}{n_{\nu}}=4.4 \times 10^{-9}$.

\begin{table}[tbp]
\centering
\begin{tabular}{|l|c|c|}
\hline
& $\nu_e$ & $\nu_{\mu, \tau}$\\
\hline 
$|\delta_\nu|$ in Water & $2 \times 10^{-8}$ & $1.3 \times 10^{-8}$\\
$|\delta_\nu|$ in $SiO_2$ (rock)& $2.5 \times 10^{-8}$ & $2.5 \times 10^{-8}$\\
$|\delta_\nu|$ in Iron & $8 \times 10^{-8}$ & $1.1 \times 10^{-7}$\\
\hline
\end{tabular}
\caption{\label{tab:deltanu} The absolute value of $\delta_\nu$ for different materials. The neutrino mass is fixed to $0.1~\text{eV}$.}
\end{table}

\begin{table}[tbp]
\centering
\begin{tabular}{|l|c|c|}
\hline
In-matter potential, $|U|$ & $\frac{G_F}{2 \sqrt{2}} ~Q_W~\rho_\text{matter}$ & $1.8 \times 10^{-14}~\text{eV}$ \\
\hline
$n-1\equiv \delta_\nu$ & $-\langle \frac{m_\nu U}{ k_\nu ^2}\rangle$& $2.5 \times10^{-8} \frac{m_\nu}{0.1~\text{eV}}$\\
\hline 
Critical angle for reflection, $\theta_\text{cr}$ & $\sqrt{-2 \delta_\nu}=\sqrt{\langle\frac{2 m_\nu U}{ k_\nu^2}\rangle}$ & $2.2 \times 10^{-4}\sqrt{\frac{m_\nu}{0.1~\text{eV}}} $\\
\hline
Critical momentum for reflection, $k_{\perp_\text{cr}}$ & $\sqrt{2 m_\nu U}$ & $6 \times 10^{-8}~\text{eV} \sqrt{\frac{m_\nu}{0.1~\text{eV}}}$\\
\hline
Extinction depth of the evanescent wave, $\lambda_\text{cr}$ & $\frac{2 \pi}{k_{\perp_\text{cr}}}$ & $3.3~\text{m} \sqrt{\frac{0.1~\text{eV}}{m_\nu}}$\\
\hline
\end{tabular}
\caption{\label{tab:parameters} Summary of relevant parameters for the $C \nu B$ and their benchmark values corresponding to the Earth, used in this paper. $Q_W$ is the weak charge of an atom as defined in eq.~\ref{eq:U}}
\end{table}

As it will become clear in section~\ref{sec:flatearth}, this overdensity extends for $\lambda_\text{cr}\equiv \frac{2 \pi}{k_{\perp_\text{cr}}}$ on either side of the Earth's surface. $\lambda_\text{cr}$ can be readily identified as the extinction depth of the evanescent neutrino wave inside the Earth and it arises through tunnelling in the classically inaccessible region. Table~\ref{tab:parameters} summarizes the quantities relevant for the refraction and reflection of neutrinos from the Earth and their characteristic size.

In what follows, we present an exact calculation of the $C\nu B$ distribution in the vicinity of the Earth's surface assuming that the Earth is flat (section~\ref{sec:flatearth}). In section~\ref{sec:roundearth}, we discuss how the Earth's curvature and surface roughness only affects our result at distances larger than $\mathcal{O}(\lambda_\text{cr})$ above and below ground. We present the expected asymmetry on the surface of the Earth in  section~\ref{sec:conclusions}, where we further discuss our assumptions and the implications of our result for the $C \nu B$ ``no-go'' theorem. For simplicity, we assume that the neutrino weak and mass eigenstates coincide.

\section{The flat Earth approximation}
\label{sec:flatearth}

\begin{figure}[tbp]
\centering 
\includegraphics[width=.8\textwidth,origin=l, trim=0 0 0 0,clip]{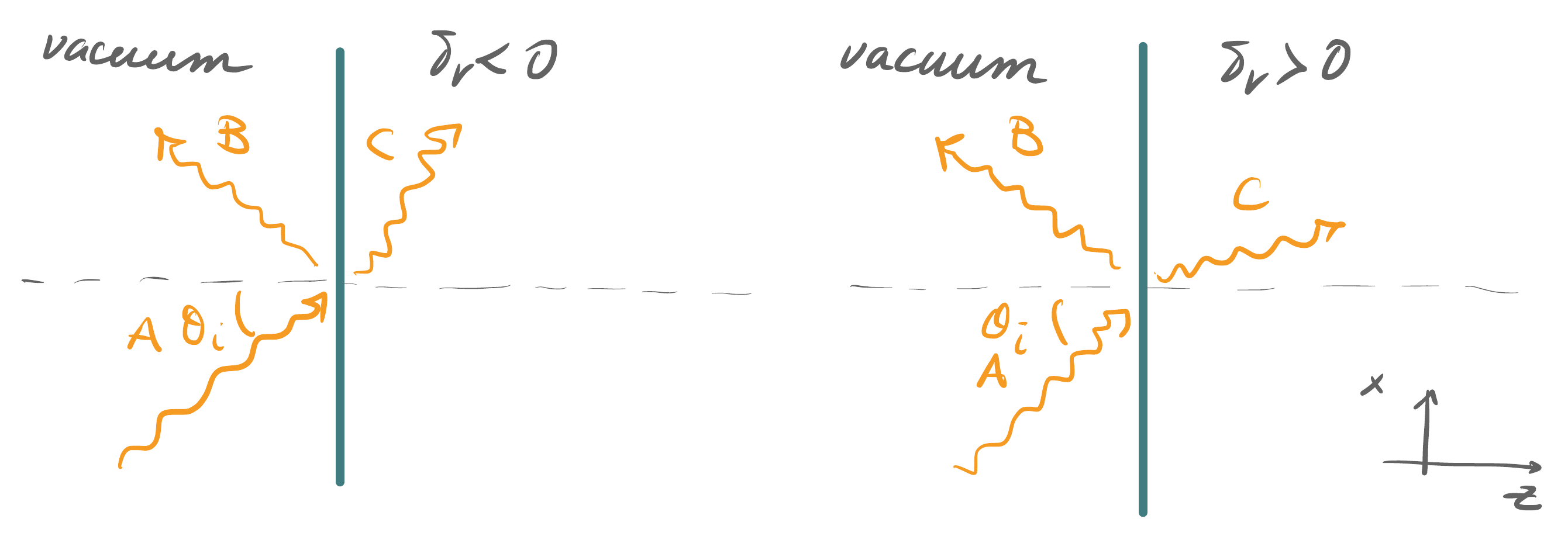}
\includegraphics[width=.9\textwidth,origin=c,angle=0]{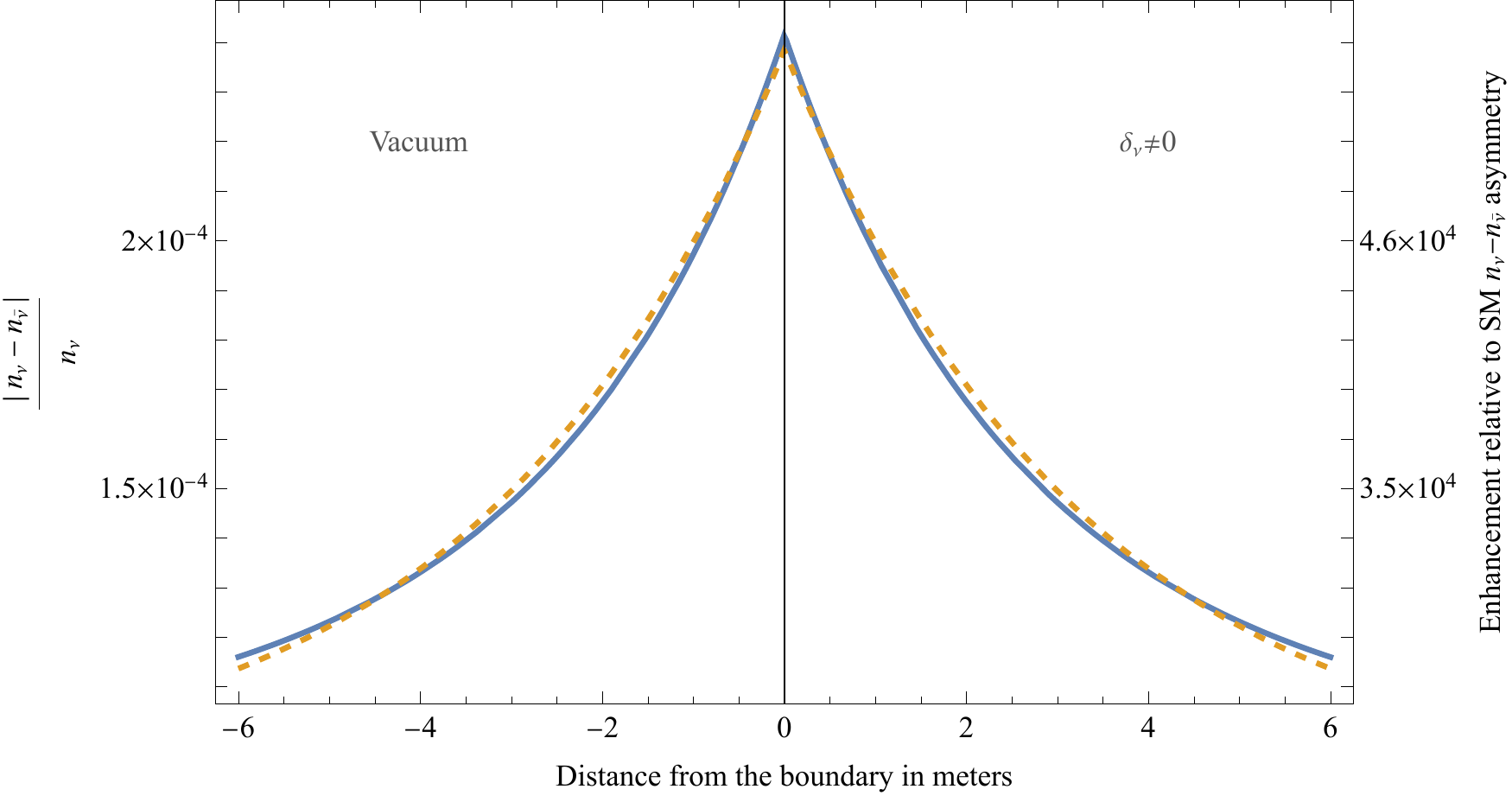}
\caption{\label{fig:boundary}  \textit{Top panel:} The flat Earth as a boundary between two media. Relic neutrinos are incident on the boundary from the left with a wave amplitude $A$. The reflected wave amplitude is $B$, and the transmitted wave $C$ is refracted at an angle larger (smaller) than the incident for $\delta_\nu <0 (\delta_\nu>0)$. \textit{Bottom panel:} The calculated fractional asymmetry corresponding to our toy model for the flat Earth shown in the top panel. The solid blue line is the exact result of eqs. \ref{eq:asymmetrywaveoutside} and \ref{eq:asymmetrywaveinside}. The dashed orange line is the approximate solution of eq.~\ref{eq:asymmetryanalyticmonochromatictotal}.}
\end{figure}

In order to calculate the distribution of cosmic neutrinos on the surface of the Earth, we will start by making the simplifying assumption that the Earth is flat and the boundary between the vacuum and a medium with index of refraction $|\delta_\nu|=2.5\times 10^{-8}$ (see figure~\ref{fig:boundary}). In this toy model, cosmic neutrinos and anti-neutrinos are hitting this interface only from the vacuum side.
We model the $C \nu B$ as an incoherent superposition of waves traveling in all directions with momentum $k_\nu\equiv \langle k_\nu^{-2} \rangle^{-1/2}=2.7 \times 10^{-4}~\text{eV}$, because this is the momentum scale implied by the all important index of refraction. We denote by $\theta_i$ the incidence angle on the vacuum-medium interface, and the directions parallel  and perpendicular to the interface are x and z, respectively, as shown in figure~\ref{fig:boundary}.

The full wave solution for each component incident on the boundary by an angle $\theta_i$ is obtained by requiring that the wave and its first derivative are continuous on the boundary. Taking the incident wave to be $\psi_\text{incident}= A e^{i(k_\parallel x+k_\perp z)}$, the reflected wave is given by:
\begin{eqnarray}
\psi_\text{reflected}= B e^{i(k_\parallel x-k_\perp z)},~\text{with} \\
B= A \frac{k_\perp -k_\perp'}{k_\perp +k_\perp '},
\end{eqnarray}
while the transmitted wave is:
\begin{eqnarray}
\label{eq:Cnus}
\psi_\text{transmitted}= C e^{i(k_\parallel x+k_\perp ' z)},~\text{with} \\
C= A \frac{ 2 k_\perp}{k_\perp +k_\perp '},
\end{eqnarray}

\begin{figure}[tbp]
\centering 
\includegraphics[width=.9\textwidth,origin=c, trim=0 0 0 0,clip]{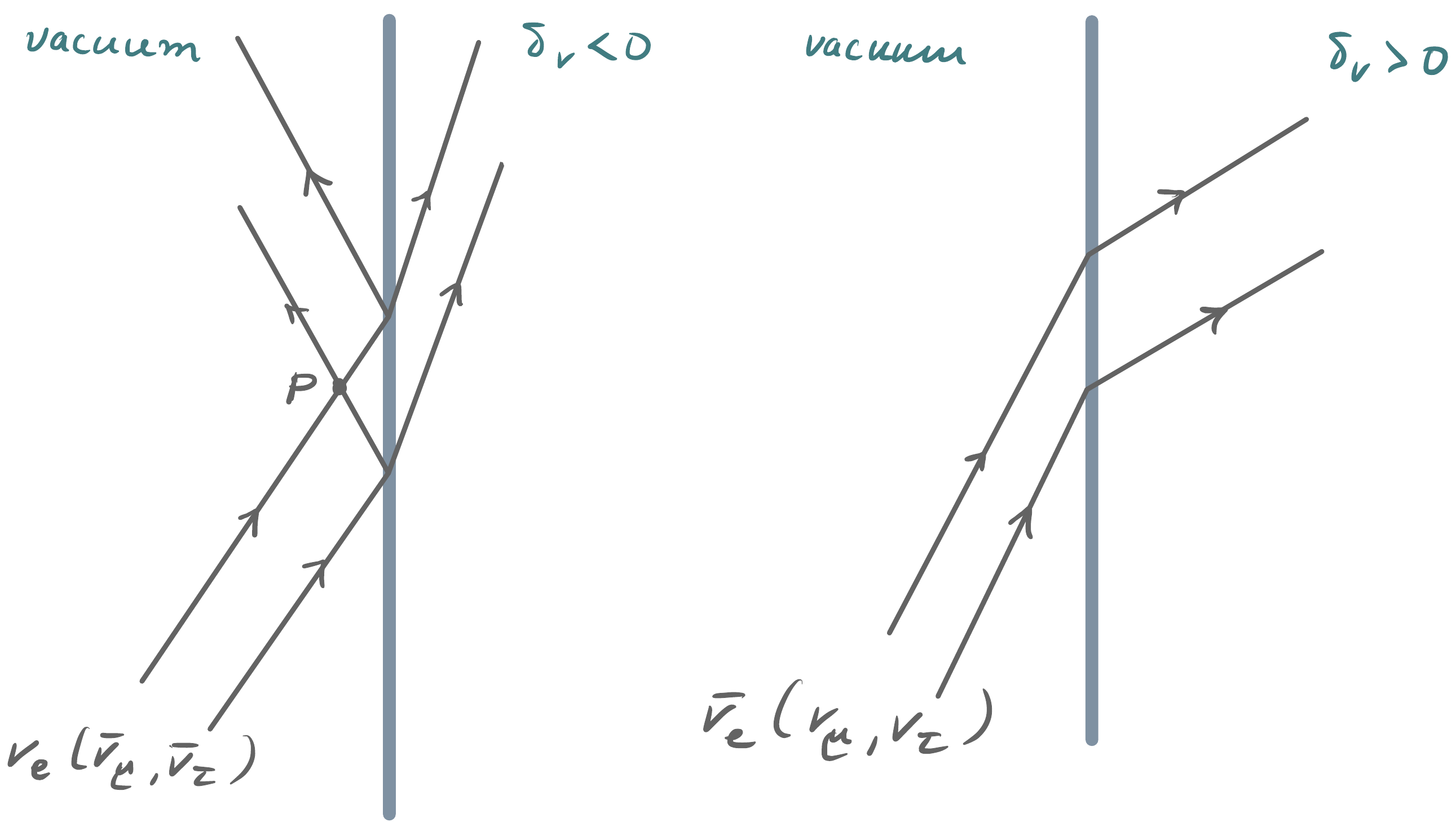}
\caption{\label{fig:flatearthrays} Geometric optics interpretation of the neutrino asymmetry.\textit{On the left:} Neutrinos with $\delta_\nu<0$ are incident on the boundary from the vacuum side. For each direction of k, each point P on the vacuum side gets intersected by two rays: one directly, and one indirectly after reflection, thereby doubling the density. Inside the medium neutrinos slow down so the rays get compressed in the z-direction, also enhancing the density. \textit{On the right:} Antineutrinos with $\delta_\nu>0$ incident on the boundary from the vacuum side can only be refracted and speed-up inside the medium. As a result, the rays fan-out and the density is diluted. }
\end{figure}

where $k_\parallel = k_\nu \sin \theta_i $, $k_\perp =  k_\nu  \cos \theta_i$, and $k_\perp ' =  k_\nu  \sqrt{\cos^2 (\theta_i) + 2 \delta_\nu}$ are the wave's momentum in the x-direction, and the z-direction in vacuum and inside the medium, respectively. Note that $k_\perp'$ becomes imaginary for negative $\delta_\nu$ and for $\cos \theta_i < \cos (\pi/2 - \theta_\text{cr})$, which signals the emergence of the evanescent wave. 

The amplitude $A$ of the incident wave is determined by the angular dependence of the $C \nu B$ density. Ignoring the relative motion of the Earth to the $C \nu B$, this distribution is angle independent, so that $\frac{\partial n_\nu}{\partial \Omega}=\frac{n_\nu}{4 \pi}$, where $\Omega$ denotes the solid angle. Given the cylindrical symmetry of our flat Earth toy model, this distribution can be simplified to $\frac{\partial n_\nu}{\partial (\cos \theta_i)}=\frac{n_\nu}{2}$. This implies  that $A=\frac{1}{\sqrt{2}}$ is the right choice so that our full wave solution directly computes the fractional changes in the neutrino density.

This fractional asymmetry on the vacuum side of the interface is calculated by integrating over the incidence angle and taking the difference between $\delta_\nu$ smaller and larger than zero:
\begin{eqnarray}
\label{eq:asymmetrywaveoutside}
\frac{n_\nu-\bar{n}_\nu}{n_\nu}\Bigg|_{z<0} =\int_0^{\pi/2} (|\psi_\text{incident}+\psi_\text{reflected}|^2_{\delta_\nu <0} -|\psi_\text{incident}+\psi_\text{reflected}|^2_{\delta_\nu >0}) d(\cos \theta_i).
\end{eqnarray}
The corresponding asymmetry inside the material is equivalently:
\begin{eqnarray}
\label{eq:asymmetrywaveinside}
\frac{n_\nu-\bar{n}_\nu}{n_\nu}\Bigg|_{z>0} =\int_0^{\pi/2} (|\psi_\text{transmitted}|^2_{\delta_\nu <0}-|\psi_\text{transmitted}|^2_{\delta_\nu >0}) d(\cos \theta_i).
\end{eqnarray}

The results of the integration are shown in figure~\ref{fig:boundary}. We see that the asymmetry takes the maximal value exactly on the boundary and over a distance set by the scale of $\lambda_\text{cr}=(3.3~\text{meters})^{-1}$ it relaxes to a non-zero value. The value of the asymmetry near the boundary is about $10^5$ times larger than the naive expectation. 

Exactly on the boundary and extinction lengths away from it, the asymmetry can be calculated analytically. Since $\delta_\nu \ll 1$, we find that:
\begin{eqnarray}
\label{eq:asymmetryanalyticmonochromatic}
\frac{n_\nu-\bar{n}_\nu}{n_\nu}\Bigg|_{z=0} =\frac{16}{15} \sqrt{2 |\delta_\nu|},~\text{and}\\
\frac{n_\nu-\bar{n}_\nu}{n_\nu}\Bigg|_{z\gg 0~\text{or}~ z\ll 0} =\frac{2}{5} \sqrt{2 |\delta_\nu|}
\end{eqnarray}
The approximate z-dependence of the asymmetry is:
\begin{eqnarray}
\label{eq:asymmetryanalyticmonochromatictotal}
\frac{n_\nu-\bar{n}_\nu}{n_\nu}(z) =\frac{2}{15} \sqrt{2 |\delta_\nu|} \left(3+5e^{-\frac{|z|}{\lambda_\text{cr}}}\right)
\end{eqnarray}

There are two surprises in this result. First,  the boundary value of the asymmetry reduces exponentially to the asymptotic value over a distance $\lambda_\text{cr}$ and \emph{not the much smaller $\lambda_{dB}$}. Second, the asymmetry persists at any distance away from the boundary. 

The first follows because the dynamics depends on the invariant momentum transfer $k_{\perp_\text{cr}}$ and not the much larger frame dependent neutrino momentum $ k_\nu $. 
We can also see this by exploiting the translational symmetry along the x-direction and going to a frame where the wave does not move parallel to the Earth. In this frame it is apparent that the important length scale of the problem is not the $C\nu B$  wavelength of $\sim mm$ but the wavelength of the neutrino perpendicular to the surface. This is of order of the wavelength of the evanescent wave, $\lambda_\text{cr}$ which is a few meters.

The second surprise -- that the asymmetry outside persists at any distance away from the boundary -- is understood by going to the geometric optics limit as  
shown in figure~\ref{fig:flatearthrays}. Geometric optics is a reliable approximation if the wave moves over distances much larger than $\lambda_\text{cr}$.
For each direction of the momentum $ k_\nu $ for which the incidence angle is more than critical, each point P on the vacuum side is intersected by two rays: one directly, and one indirectly after total reflection. This doubles the $\nu_e$ density at P coming from such directions.
Therefore the reflected rays are responsible for the increase in the density of $\nu_e$s. Since there are no reflected rays for the electron antineutrinos, this results in an electron neutrino-antineutrino asymmetry. 

Inside the material, the value of the asymmetry can be understood due to the slowdown of neutrinos for which the medium generates a repulsive potential. So, even though some of the rays of $\delta_\nu <0$ neutrinos are indeed lost due to reflection, the rays that refract to angles close to but above critical, get compressed relative to those for $\delta_\nu >0$ neutrinos, as drawn in figure~\ref{fig:flatearthrays}. The fact that the asymptotic value of the asymmetry is the same both inside and outside is only true for refractive indices very close to 1 and changes quantitatively and qualitatively in the more general case.

So far we have treated the neutrino background as monochromatic with momentum $ k_\nu $. In order to take into account the momentum distribution of neutrinos, we perform an appropriate averaging that includes the Fermi-Dirac distribution of neutrinos, i.e.:
\begin{eqnarray}
\frac{n_\nu-\bar{n}_\nu}{n_\nu}=\frac{\int \frac{n_\nu-\bar{n}_\nu}{n_\nu}(k) \frac{k^2}{e^{k/T_\nu}+1} dk }{\int \frac{k^2}{e^{k/T_\nu}+1} dk},
\end{eqnarray}
where $\frac{n_\nu-\bar{n}_\nu}{n_\nu}(k)$ is given by~eqs.~\ref{eq:asymmetryanalyticmonochromatic} taking into account that $\delta_\nu \propto k^{-2}$. The value of the extinction depth $\lambda_\text{cr}$ is independent of the momentum. This allows us to write down an analytic form of the excess around the boundary that includes the effects of the effects of color aberrations in the reflection and refraction of neutrinos from the boundary:
\begin{eqnarray}
\label{eq:asymmetryexact}
\frac{n_\nu-\bar{n}_\nu}{n_\nu}(z)=0.73 \frac{2}{15}\sqrt{2|\delta_\nu|}\left(3+5e^{-\frac{|z|}{\lambda_\text{cr}}}\right)
\end{eqnarray}
Here $|\delta_\nu|=2.5 \times 10^{-8}$, is the benchmark value of the index of refraction in the Earth for neutrinos of momentum $ k_\nu$. Taking into account the full Fermi-Dirac distribution is thus only a small correction.

To summarize, in our toy model of a flat Earth, we find an excess of neutrinos with $\delta_\nu<0$ around the surface of the Earth that is enhanced by several orders of magnitude relative to the SM expectation. The excess persists away from the boundary, both in the interior and the exterior of the Earth.

\begin{figure}[tbp]
\centering 
\includegraphics[width=.9\textwidth,origin=c, trim=0 0 0 0,clip]{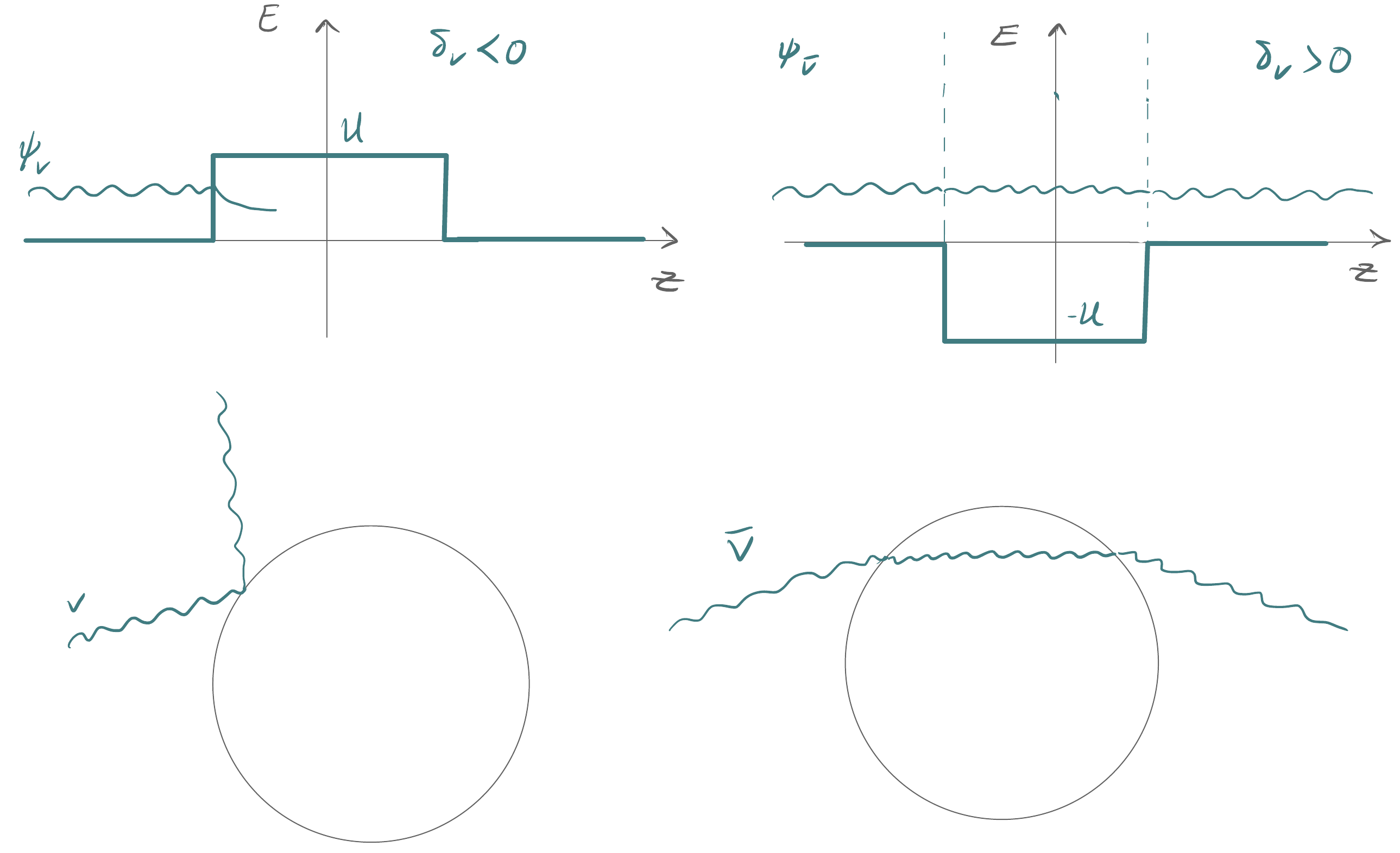}
\caption{\label{fig:rays-dec-acc} \textit{Top:} On the left, neutrinos reflecting from a repulsive potential $U$ in 1-D. On the right, antineutrinos are refracted from an attractive potential $U$. Antineutrinos that are attracted move faster in an attractive $U$, while neutrinos are reflected and cannot access classically the region of a repulsive $U$.
 \textit{Bottom:} The 3-D analog of the same problem as above. For antineutrinos, entering the region of $U$ includes refraction following Snell's law.}
\end{figure}

One may wonder why in our toy model we assumed all particles to be incident from the vacuum side. After all, antineutrinos \footnote{Neutrino and antineutrino from now on will refer to $\nu_e (\bar{\nu}_{\mu,\tau})$ and $\bar{\nu}_e (\nu_{\mu,\tau})$, respectively.} incident from the interior of the Earth could behave just like the neutrinos incident from the outside and create an excess that cancels the excess we just computed. The reason that this does not happen 
is that in the realistic case both neutrinos and antineutrinos originate at infinity and enter the Earth from the outside. This is shown schematically in figure~\ref{fig:rays-dec-acc} for both the one dimensional and three dimensional cases. In the one 1-D case we see that $\nu s$ with sub-critical momentum attempting to penetrate the 1-D barrier pile-up and create a $\nu$ excess on the surface. In contrast, $\bar{\nu} s$ of the same energy do not encounter any barrier and ``fly over'' the potential, speed-up and do not create a $\bar{\nu}$ excess neither at their entry nor at their exit points.

Similarly in the 3-D case, neutrinos attempting to enter
the Earth from the outside at super-critical angles get totally reflected and create a pile-up excess on the surface.
In contrast, antineutrinos entering the sphere do not undergo total internal reflection in the sphere.
They just transverse the sphere and, by time-reversal and spherical symmetry, leave at the same angle as they entered.
 Thus they do not pile up or generate an evanescent wave that enhances their number density near the Earth's surface. 
 So, the spherical Earth acts like a deflector of antineutrinos that come from the outside and prevents them from ever hitting its surface from its interior at a supercritical angle, thereby avoiding a pile-up.
 We will return to this topic in the next section.

\section{The Round Earth}
\label{sec:roundearth}

Alas, the Earth is not flat. 
How far from the surface can we trust the flat Earth approximation? To answer this first consider a point P outside the Earth and at a distance much larger than $\lambda_\text{cr}$, where we can trust the geometric optics approximation and treat neutrinos as rays. We assume that the Earth is a perfect sphere of size $R_\oplus=6371~\text{kms}$. 
As explained in the previous section, for an infinite, flat Earth, the enhancement of the $\nu-\bar{\nu}$ asymmetry arises because each point P (see figure~\ref{fig:flatearthrays}), no matter how far, has two parallel rays contributing to the asymmetry: one that encounters P before hitting the Earth's surface, and another that encounters P after it undergoes total reflection on the Earth's surface. The fraction of rays for which this doubling occurs, comes within a solid angle $\mathcal{O}(\theta_\text{cr})$.

\begin{figure}[tbp]
\centering 
\includegraphics[width=.48\textwidth,origin=c,angle=0]{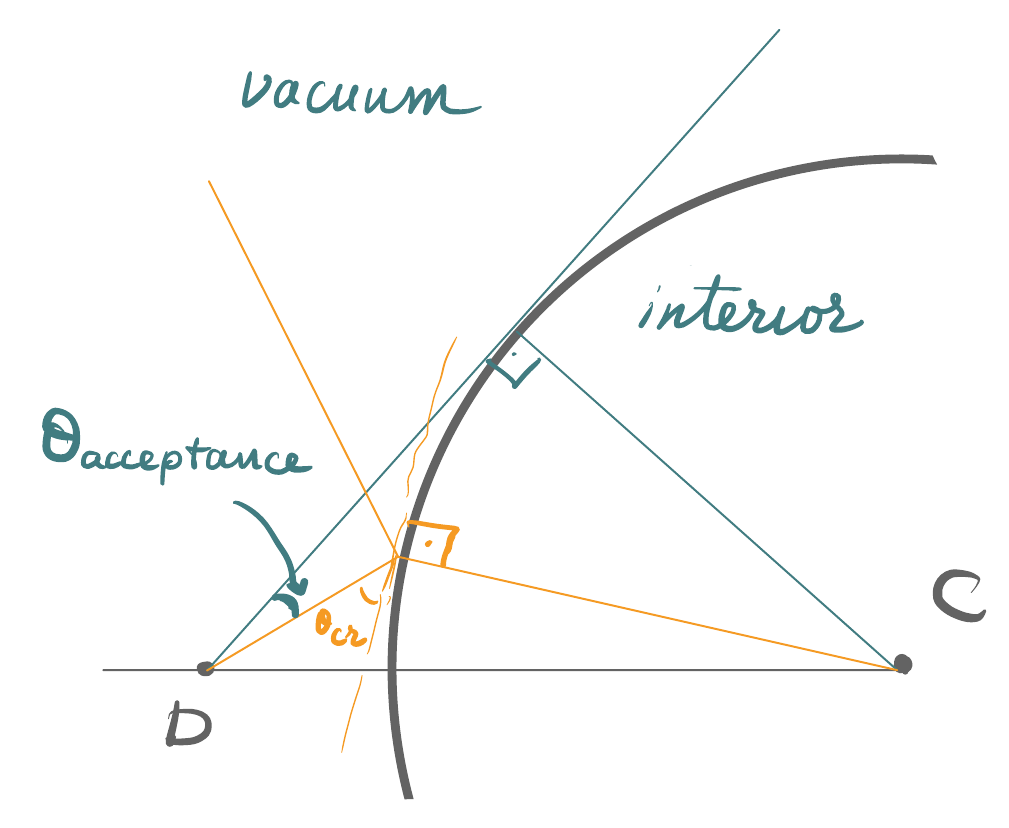}
\hfill
\includegraphics[width=.42\textwidth,origin=c, trim=0 0 0 0,clip]{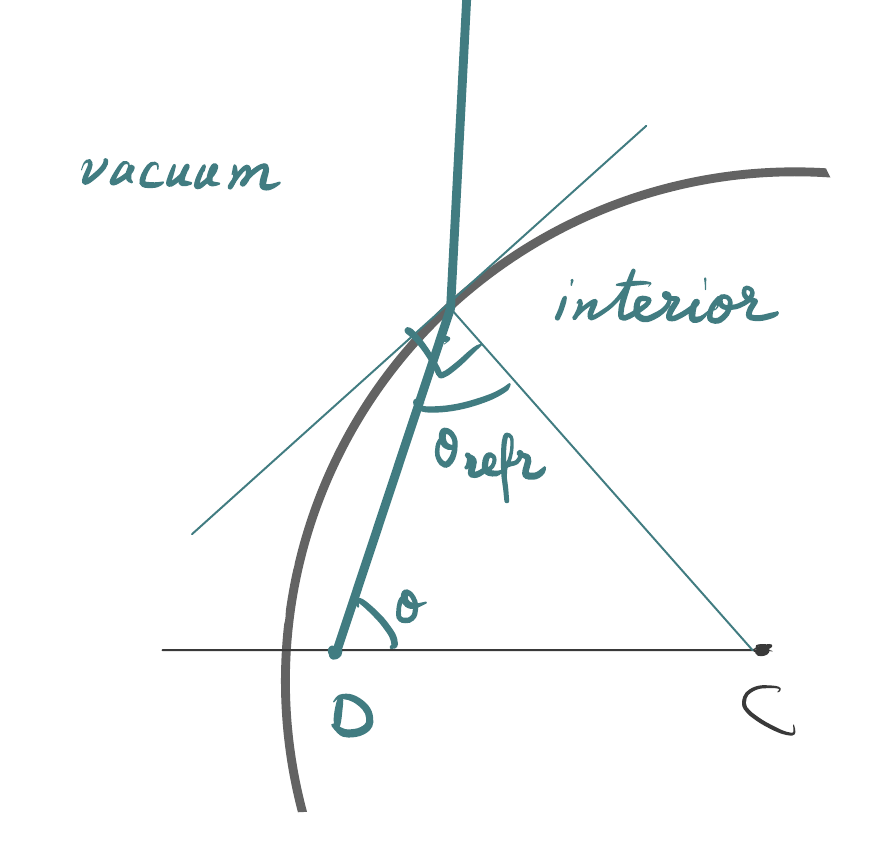}
\caption{\label{fig:roundearthrays}  \textit{On the left:}  Two rays incident on the detector D after being reflected from the Earth's surface. Only rays coming within $\theta_{\text{acceptance}}$ contribute to the asymmetry. $\theta_{\text{acceptance}}$ goes quickly to zero as D moves away from the surface. \textit{On the right:} A ray incident on the surface of the sphere arrives at the position of the detector D from an angle $\theta$ after being refracted at $\theta_{\text{refr}}$. The detector is in the Earth's interior. There is no $\theta$ for which $\theta_{\text{refr}}$ is of order the critical angle, unless the detector is close to the surface(see text).}
\end{figure}

For a curved Earth, one can still find two parallel rays that intersect point P, one before hitting the Earth's surface, and a reflected one. What changes now is that the solid angle for which this doubling occurs goes to zero as the point P moves away from the Earth's surface. This is shown pictorially in figure~\ref{fig:roundearthrays}. The solid angle is of size $\mathcal{O}(\theta_\text{cr})$ only if the point P is $\mathcal{O}(\theta_\text{cr}^2 R_\oplus)$ from the surface of the Earth. Numerically, $\theta_\text{cr}^2 R_\oplus \sim 30~\text{cm}$ which is smaller than $\lambda_\text{cr}$. Since this geometric optics argument is only valid at distances much larger $\lambda_\text{cr}$, we conclude that the enhancement of the asymmetry will not extend  at heights much larger than $\lambda_\text{cr}$.

Now let's consider a point P in the interior of the Earth. In the previous section, we saw that the enhancement of the $\nu-\bar{\nu}$ asymmetry persisted in the interior due to the significant slow down of neutrino rays refracting at angles above but close to critical (see figure~\ref{fig:flatearthrays}). If this point P is now located inside the Earth, it can only be intersected by rays that refracted to angles close to critical when it is  $\mathcal{O}(\theta_\text{cr}^2 R_\oplus)$ from the Earth's surface (see figure~\ref{fig:roundearthrays}). So, the $\nu-\bar{\nu}$ asymmetry 
will not persist at depths much larger than $\lambda_\text{cr}$ below the Earth's surface. 

In conclusion, both outside and inside the Earth, at  heights or depths of order $\lambda_\text{cr}$ when the flat Earth approximation is valid we find a significant enhancement of the $\nu-\bar{\nu}$ asymmetry, whereas at heights/depths much larger than 
$\lambda_\text{cr}$ when geometric optics is valid we find no such enhancement.

 A realistic extension of our flat Earth computation will, in addition to the Earth's curvature, include the profile of the local terrain together with the effects of the roughness on the Earth's surface reflectivity. This latter issue has been studied extensively in the context of reflection of light from surfaces~\cite{Bennett:61}. It has been shown that, as long as the surface is much larger than $\lambda_\text{cr}$, the reflectivity of an imperfect surface is related to the reflectivity of a perfectly flat surface by a factor of $e^{-({k_{\perp}} \sigma)^2}$, where $\sigma$ is the rms value of the surface's uneveness. This suggests that, if there is a patch of the Earth's surface that is much larger than $(3.3~\text{meters})^2$ with ``bumps'' much smaller than $\lambda_\text{cr}$, then our flat Earth result should hold at least within $3.3~\text{meters}$ above and below the surface.

\begin{figure}[tbp]
\centering 
\includegraphics[width=.9\textwidth,origin=c, trim=0 0 0 0,clip]{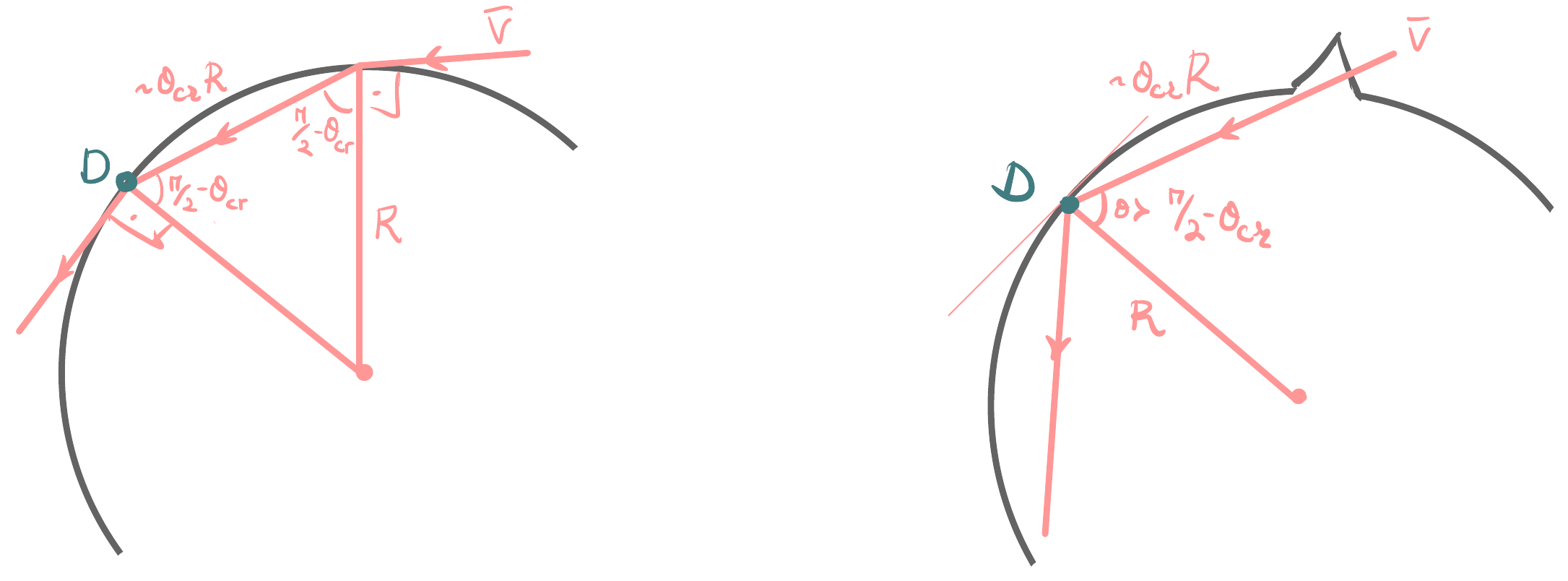}
\caption{\label{fig:antinusearth} \textit{On the left:} For a perfectly spherical Earth, the antineutrinos entering the Earth and hit the detector at D from below the surface can only be refracted at a maximum angle of $\frac{\pi}{2}-\theta_\text{cr}$, so there is essentially no possibility of reflection at the position of the detector. This maximally refracted $\bar{\nu}$s come from a distance of 
$\mathcal{O}(\theta_\text{cr} R_\oplus)\sim 1~\text{km}$
 \textit{On the right:} For an imperfect Earth, the presence of a mountain can allow for antineutrino rays to enter the Earth and hit D at an angle larger than critical, so reflection is now possible.}
\end{figure}

We now continue our discussion of antineutrinos coming from under the surface of the Earth, that we started at the end of section~\ref{sec:flatearth}. These antineutrinos could in principle develop an excess reducing the neutrino excess.
However this does not happen. For antineutrinos to hit a detector at point D, as shown in figure~\ref{fig:antinusearth}, they first must enter the Earth from the outside. For a perfectly spherical Earth, the entering antineutrinos can be refracted to a maximum angle of $\frac{\pi}{2}-\theta_\text{cr}$; this is not enough to give total reflection from the Earth's interior and create a antineutrino density. This is the geometric analog of the antineutrino ``fly-over'' argument that was presented at the end of section~\ref{sec:flatearth}.
A spherical Earth acts like a filter redirecting antineutrinos that would have totally reflected from under the Earth's surface. This eliminates the possibility of an anti-neutrino excess reducing the neutrino excess.

The presence of an imperfection such as a mountain can change this picture (see right panel of figure~\ref{fig:antinusearth}). Antineutrinos can now enter the Earth from the mountain's side and hit D at an angle larger than critical. The maximum distance for which this occurs is again $\mathcal{O}(\theta_\text{cr} R_\oplus)\sim 1~\text{km}$, otherwise totally reflecting from the interior of the curved surface  is impossible.  
This implies that as long as there is a perfectly spherical patch of Earth that is much larger than $1-10~\text{km}^2$, ignoring the antineutrinos is a good assumption. We know that the Earth has spots with height variations of less than $\lambda_\text{cr} \sim 3~m $ that are hundreds of thousands $\text{km}^2$ in size, so this is not a strong requirement. This flat spot is also small enough for us to ignore the Earth's ellipticity -- its effect is to modify angles by less than $\sim10^{-6} \ll \theta_\text{cr}$.

The argument above ignores antineutrinos that (1) are coming from the far side of the Earth and have been randomly refracted by the inhomogeneities of the Earth or (2) have been ``guided'' to the detector at the center of the flat spot by continuous reflections along the curved surface of the Earth after entering the interior by distant inhomogeneities. In case (1), the antineutrino momentum cannot be significantly altered in the Earth's interior, since the fractional average momentum change is of order $|\delta_\nu|$. In scenario (2), by analogy to guided optical modes propagating in curved fibers~\cite{marcuse1976curvature}, the antineutrinos cannot survive multiple reflections, they tunnel outside and leave the Earth. 

So both (1) and (2) are too small to significantly affect the neutrino asymmetry.
Furthermore, it would be an accident if the deviations of the Earth from perfect homogeneous sphere conspire to produce a distribution
of antineutrinos inside that precisely cancels the contribution of neutrinos incident from outside the surface.

\section{Forces, Torques, and Evading no-gos}
\label{sec:forcesandtorques}

We have pointed out two new effects on the $C \nu B$ near the Earth. One is the enhancement of the $\nu-\bar{\nu}$ asymmetry, and the other is the distortion of the neutrino profile that gives rise to a gradient of the density. Both of these can impact searches for the $C \nu B$. The asymmetry enhancement increases the magnitude of torques and forces. More importantly, the density gradient gives rise to forces between the $C \nu B$ and matter that are first order in $G_F$. This evades the no go theorem of ~\cite{Cabibbo:1982bb,Langacker:1982ih} from 1982 on the absence of such forces, which discouraged searches for mechanical effects of the $C \nu B$. Next we briefly discuss two classes of experiments. 

\subsection{The Stodolsky Effect}

This effect ~\cite{Stodolsky:1974aq} does not depend on neutrino density gradients and therefore survives the ``no-go'' theorem of ~\cite{Cabibbo:1982bb,Langacker:1982ih}.  A spin that moves relative to the $C \nu B$ with velocity $\upsilon_{rel}$, experiences an energy shift of size $\Delta E= \frac{G_F}{2 \sqrt{2}}(n_\nu-n_{\bar{\nu}}) \vec{\upsilon}_{rel}\cdot \vec \sigma$. This energy splitting arises through the weak vector-axial exchange between neutrinos and matter. Given the measured baryon asymmetry, the expectation for this energy splitting predicted in ~\cite{Stodolsky:1974aq} is $10^{-47}~eV$. This energy splitting will cause spin precession, and will provide a (miniscule) signal in an NMR type experiment. If it is electron spins interacting with the $C\nu B$, this will produce a mechanical torque on an iron ferromagnet $\sim10^{-40}~N\cdot m \frac{V}{(10~\text{cm})^3}$, where V is the volume, as also predicted in the original paper~\cite{Stodolsky:1974aq}.   

The Earth's presence as discussed in this paper, enhances locally the $\nu-\bar{\nu}$ asymmetry and boosts the Stodolsky effect to a size of $\Delta E=4.6\times 10^{-43}~eV$ for a single spin, or equivalently $6\times 10^{-36}~N\cdot m \frac{V}{(10~\text{cm})^3}$ for the mechanical torque, using the benchmark values of table~\ref{tab:parameters}.

\subsection{$\mathcal{O}(G_F)$ Force and Torsion Balance Searches}
The vector-axial interaction responsible for the Stodolsky effect is subdominant to the vector-vector interaction that is also responsible for the index of refraction presented earlier. The vector-vector interaction of the $C \nu B$ with an atom is enhanced by a factor of order the atomic number and it is not suppressed by the relative velocity of the neutrino relic sea compared to the vector-axial one. The interaction potential, $U$, of eq.~\ref{eq:U} is a manifestation of the matter effect on $C \nu B$. Newton's third law suggests that there should be an analogous effect on an atom from the $C\nu B$:

\begin{eqnarray}
\label{eq:Umatter}
U_{C \nu B}=\frac{G_F}{2 \sqrt{2}}\times 
\begin{cases}
    (3Z-A)(n_{\nu_e}-n_{\bar{\nu}_e})\\
    (Z-A)(n_{\nu_{\mu,\tau}}-n_{\bar{\nu}_{\mu,\tau}}).
\end{cases}
\end{eqnarray}

This atom can thus experience a force:

\begin{eqnarray}
\label{eq:forcematter}
\vec{F}_{C \nu B}=- \vec{\nabla} U_{C \nu B} \propto \vec{\nabla}(n_{\nu}-n_{\bar{\nu}})
\end{eqnarray}

If the $C \nu B$ density is uniform, then the force is zero -- this is the argument made explicitly in~\cite{Cabibbo:1982bb}, and the essence of the non-trivial proof presented in~\cite{Langacker:1982ih} for the ``no-go'' theorem on $\mathcal{O}(G_F)$ forces from the $C \nu B$.

As we presented above, the reflection of neutrinos from the Earth's surface produces a local $C \nu B$ density gradient, so the force on matter is no longer zero. On a $10~\text{cm}$ block of Tungsten   ($^{184}_{74} W$)this force is $1.6\times 10^{-31}~N\frac{V_{\text{Tungsten}}}{(10~\text{cm})^3}$. If the same amount of Tungsten is incorporated in a torsion balance set-up the interaction with the $C \nu B$ appears as an equivalence principle violating torque of size:
\begin{eqnarray}
\tau_{C \nu B}=1.6 \times 10^{-32} N\cdot m \frac{V_{\text{Tungsten}}}{(10~\text{cm})^3} \frac{\ell_\text{torsion}}{10~\text{cm}}, 
\end{eqnarray}

where $\ell_{\text{torsion}}$ is the effective arm-length of the torsion balance, using again the parameters of table~\ref{tab:parameters}. 

The presence of the Earth enhances the local $\nu-\bar{\nu}$ asymmetry and creates a gradient giving rise of a non-zero $\mathcal{O}(G_F)$ force that can produce a torque that is a factor of $10^3$ larger than the corresponding Stodolsky effect. The torque due to the Earth-induced gradient force ultimately boosts the size of $\mathcal{O}(G_F)$ effects on matter by a factor of  $10^8$, when compared to the $10^{-40} N\cdot m$ size Stodolsky torque based on what was believed so far.

Both Earth-induced refractive effects estimated above, while orders of magnitude away from what has currently being achieved in a laboratory setting, are orders of magnitude larger than previously thought.

\section{Conclusions, Comments, and Summary}
\label{sec:conclusions}

\begin{figure}[tbp]
\centering 
\includegraphics[width=.9\textwidth,origin=c, trim=0 0 0 0,clip]{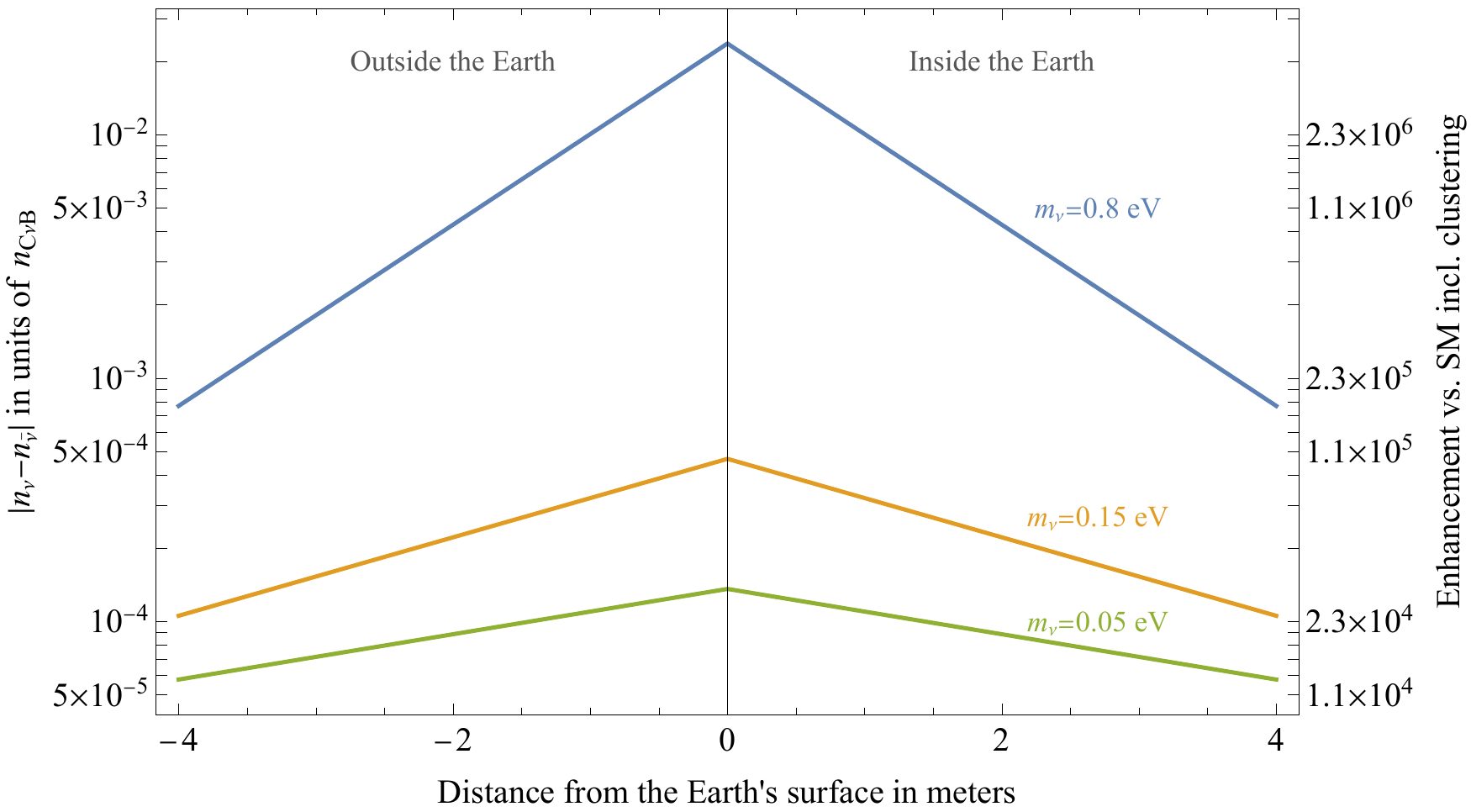}
\caption{\label{fig:earthasymmetry} The neutrino asymmetry of neutrinos and antineutrinos normalized to the $C\nu B$ density of one neutrino helicity, $n_{C \nu B}=56~\text{cm}^{-3}$, as a function of the distance from the Earth's surface for three different neutrino masses, $m_\nu=0.8$~eV (top curve), $m_\nu=0.15$~eV (middle curve), and $m_\nu=0.05$~eV (bottom curve). We have included enhancements in the density due to possible local clustering of neutrinos, as estimated in~\cite{Alvey:2021xmq}.}
\end{figure}

The main results of this work are summarized in figure~\ref{fig:earthasymmetry} which shows the expected absolute size of the neutrino asymmetry as a function of the distance from the surface of the Earth for three different neutrino masses: 0.8, 0.15,~and 0.05~eV. The value of 0.8~eV is consistent with the KATRIN result, but inconsistent with cosmological bounds, so we take this as an optimistic upper bound. 0.15~eV is closer to the cosmological constraint~\cite{Esteban:2020cvm}, while we take 0.05~eV as a lower bound on the heaviest neutrino mass~\cite{Esteban:2020cvm}. Figure~\ref{fig:earthasymmetry} includes chromatic effects, as calculated in eq.~\ref{eq:asymmetryexact} and assuming that the asymptotic asymmetry is the SM expectation, as well as possible enhancements to the local $C\nu B$ density due to clustering~\cite{Ringwald:2004np,deSalas:2017wtt,Zhang:2017ljh,Mertsch:2019qjv}. The clustering factor has been estimated from simulations to be $77\left(\frac{m_\nu}{1~\text{eV}}\right)^{2.2}$~\cite{Alvey:2021xmq}. Clustering can be relevant for neutrinos of 0.8 and 0.15~eV in mass, but becomes unimportant when the heaviest neutrino mass is 0.05~eV. In the most optimistic scenario, the local $\nu_e$ and $\bar{\nu}_{\mu,\tau}$ excess can be up to $10^{-2}$ of the relic neutrino density per helicity state of $56~\text{cm}^{-3}$ -- over $10^6$ times larger than the naive expectation. We should state that the results of figure~\ref{fig:earthasymmetry} are an approximation since they do not take into account the local terrain shape and composition effects, as well as any changes in the neutrino phase space distribution due to clustering. Our analysis also ignores the effects of Pauli's exclusion principle, which we do not expect to be significant.

To simplify our calculations, we have taken the neutrino mass and weak eigenstates to coincide. In the real world, the $C \nu B$ neutrinos today have each collapsed to a mass eigenstate that is a linear combination of different weak eigenstates. Barring accidental cancellations between the weak charges and mixing angles of these linear combinations,  we expect similar effects for all neutrinos that have masses in a similar range of $0.1~\text{eV}$ considered here. We also stress that the results presented here strictly apply to Dirac neutrinos. We will present the Majorana case in an upcoming paper. 

Since the heaviest neutrino has to be at least $0.05~\text{eV}$, there is at least one neutrino exhibiting the effects pointed out here. Even for lighter neutrinos there are similar effects, with the asymmetry reduced by just a factor of $\sqrt{\frac{m_\nu}{0.1~\text{eV}}}$ and the shell size enhanced by $\sqrt{\frac{0.1~\text{eV}}{m_\nu}}$.

 In conclusion, in this paper we argue that total reflection of cosmic neutrinos from the surface of the Earth results in a local $\nu-\bar{\nu}$ asymmetry, that far exceeds the one implied by the measured baryon asymmetry. It also produces a gradient of the neutrino density that evades the forty-year-old ``no-go'' theorems on the vanishing of $\mathcal{O}(G_F)$ forces. Total reflection overcomes the handicap of weak interactions as it scales like $\sqrt{G_F}$, contrary to refraction which is $\propto G_F$, or scattering which has the even less favorable scaling of $G_F^2$. Notably, the $\mathcal{O}(G_F)$ force due to the presence of the Earth is now larger than the one produced by coherent scattering \cite{Shvartsman:1982sn, gelmininussinov:2001hd, Domcke:2017aqj, Shergold:2021evs, Bauer:2022lri}.

 Our arguments are based on matching the exact wave calculation presented in section~\ref{sec:flatearth}, and the geometric optics arguments of section~\ref{sec:roundearth}. 
There may well be much to be gained by improving these treatments.


\acknowledgments

We thank Giorgio Gratta and Junwu Huang for very valuable discussions. We also thank  John March-Russell, Matheus Hostert, Neal Dalal, Tom Abel, Maxim Pospelov, Masha Baryakhtar, Sebastian Baum, and Davide Racco, for useful conversations.  
AA is grateful for the support of the Stavros Niarchos Foundation and the Gordon and Betty Moore foundation.
SD is grateful for support from the National Science Foundation under Grant No. PHYS-
2014215, and from the Gordon and Betty Moore Foundation Grant GBMF7946. Research
at Perimeter Institute is supported in part by the Government of Canada through the Department of Innovation, Science and Economic Development Canada and by the Province
of Ontario through the Ministry of Colleges and Universities.

\bibliography{refs}

\end{document}